\documentclass[aps,twocolumn,prd,showpacs,nofootinbib]{revtex4}
\usepackage{amsmath}
\usepackage{graphicx}
\usepackage{dcolumn}
\usepackage{bm}
\usepackage{amssymb}
\usepackage{latexsym}

\def\be{\begin{equation}}
\def\ee{\end{equation}}
\def\ba{\begin{eqnarray}}
\def\ea{\end{eqnarray}}

\newcommand{\beqa}{\begin{eqnarray}}
\newcommand{\eeqa}{\end{eqnarray}}
\newcommand{\bea}{\begin{eqnarray}}
\newcommand{\eea}{\end{eqnarray}}

\newcommand{\vp}{\varphi}

\def\beq{\begin{equation}}
\def\eeq{\end{equation}}

\begin{document}

\title{Inflation in string-inspired cosmology
       and suppression of CMB low multipoles}

\author{
Yun-Song Piao$^{1,2}$, Shinji Tsujikawa$^{3}$ and Xinmin
Zhang$^{1}$}

\affiliation{${}^1$Institute of High Energy Physics, Chinese
Academy of Science, P.O. Box 918-4, Beijing 100039, P. R. China}
\affiliation{${}^2$Interdisciplinary Center of Theoretical
Studies, Chinese Academy of Sciences, P.O. Box 2735, Beijing
100080, China} \affiliation{${}^3$Institute of Cosmology and
Gravitation, University of Portsmouth, Mercantile House,
Portsmouth PO1 2EG, UK}

\date{\today}

\begin{abstract}

We construct a string-inspired nonsingular cosmological scenario
in which an inflaton field is driven up the potential before the
graceful exit by employing a low-energy string effective action
with an orbifold compactification. This sets up an initial
condition for the inflaton to lead to a sufficient amount of
$e$-foldings and to generate a nearly scale-invariant primordial
density perturbation during slow-roll inflation. Our
scenario provides an interesting possibility to explain a
suppressed power spectrum at low multipoles due to the presence of
the modulus-driven phase prior to slow-roll inflation 
and thus can leave a strong imprint of 
extra dimensions in observed CMB anisotropies.

\end{abstract}

\vskip 1pc  \vskip 2pc
\maketitle \baselineskip = 12pt

String theory has been regarded as one of the promising candidate
theories of quantum gravity over the past 20 years \cite{string}.
However it is unfortunately not easy to detect the signature of
extra dimensions and any stringy effect in accelerators in the
foreseeable future. In this sense string cosmology presumably
provides the only means of testing string or M-theory concretely.
It is therefore very important to investigate the viability of
string theory by exploring possible cosmological implications.

From the observational side, the recent WMAP measurements showed a
lack of power on the largest scales and the running of the
spectral index in temperature anisotropies \cite{Spergel}. This is
difficult to be explained by standard slow-roll inflation,
thus may have roots in esoteric Planck scale physics. One can
consider an exciting possibility where the suppression of power
originates from some stringy effect which may be present prior to
standard inflation (see e.g., Refs.~\cite{sup} for a number of
attempts to explain this suppression).

String theory provides an interesting possibility to avoid the
big-bang singularity due to its underlying symmetries \cite{BV}.
The construction of nonsingular cosmological models based on
the low-energy string effective action was pioneered by Veneziano
and Gasperini \cite{Veneziano:91}, whose model is called the
Pre-Big Bang (PBB) scenario. According to this scenario, the
universe exhibits a super-inflation when the dilaton field, which
characterizes the strength of the string coupling parameter,
evolves toward the strong coupling regime. The curvature
singularity can be avoided by implementing the dilatonic loop and
higher-derivative corrections to the tree-level action, which
is followed by a post-big bang phase with a decreasing Hubble rate
\cite{Bru}.

While the PBB scenario is an intriguing attempt to merge string
theory and cosmology, the spectrum of curvature perturbations is
highly blue-tilted ($n\simeq 4$) in its simplest form
\cite{BGGMV}, which is incompatible with observations
\cite{Spergel}. If we take into account the contribution of an
axion field, the spectrum of the axion perturbation can be
scale-invariant depending on the expansion rate of an internal
dimension \cite{Copeland:97}. Although there remains a possibility
to explain the observationally supported flat spectra if the axion
plays the role of the curvaton \cite{curvaton}, the PBB scenario
still suffers from the problem of the requirement of an
exponentially large homogenous region in order to solve the major
cosmological problems \cite{Kaloper}.

If standard slow-roll inflation occurs after a graceful exit,
this solves the homogeneity problem above as long as we have
sufficient amount of $e$-foldings ($N \gtrsim 60$). In this paper
we shall investigate a scenario in which the PBB phase is followed
by slow-roll inflation with an assumption that there exists a
light scalar field (inflaton) other than the dilaton\footnote{
This field can be regarded as one of the modulus fields which is
always present in string theory.}. Note that a similar idea was
proposed in Ref.~\cite{Piao} where slow-roll inflation is
preceded by a contracting phase. In this work we implement the
tree-level $\alpha'$ Gauss-Bonnet correction to the string
effective action in order to construct viable nonsingular
solutions. We will show that the inflaton can be driven up the
potential hill due to a nontrivial background dynamics prior to 
the graceful exit, thereby setting up sufficient initial conditions
for following slow-roll inflation. This has a similarity to
the inflation in loop quantum gravity \cite{quantum}, but the
observational signatures are different. In fact we shall see that
the presence of the kinematically driven super-inflation stage before
the graceful exit can lead to a stronger suppression of the power
spectrum on largest scales, as favoured from observations.

We start with the following 4-dimensional action that appears for
the orbifold compactification in low-energy effective string
theory \cite{Antoniadis}
\begin{eqnarray}
 S = \int d^4 x \sqrt{-g} \left[ \frac{R}{2}
 - \frac14 (\nabla \phi)^2
 - \frac34 (\nabla \sigma)^2
 +{\cal L}_{\rm inf}
 +{\cal L}_c
 \right]\!,
\label{Bein}
\end{eqnarray}
which is written in the Einstein frame. Here $R$ is the scalar
curvature and the fields $\phi$ and $\sigma$ correspond to the
dilaton and the modulus, respectively. We shall consider a
minimally coupled inflaton field, $\chi$, with lagrangian
\begin{eqnarray}
 {\cal L}_{\rm inf}=-(1/2)(\nabla \chi)^2-V(\chi)\,.
\label{Linf}
\end{eqnarray}
Hereafter we assume that the inflaton has a mass $m_{\chi}$ with
potential $V(\chi)=(1/2)m_{\chi}^2\chi^2$.

The correction term ${\cal L}_c$ is given by
\begin{eqnarray}
 {\cal L}_c=(1/16) \left[\lambda
 e^{-\phi}-\delta\xi(\sigma) \right] R_{\rm GB}^2\,,
\label{Lc}
\end{eqnarray}
where $R_{\rm GB}^2 =R^2-4R^{\mu\nu}R_{\mu\nu}+
R^{\mu\nu\alpha\beta}R_{\mu\nu\alpha\beta}$ is a Gauss-Bonnet
term. Note that we are considering the correction up to the first
order in $\alpha'$. The coefficients, $\lambda$ and $\delta$, are
determined by the inverse string tension $\alpha'$ and the
4-dimensional trace anomaly of the $N=2$ sector. $\lambda$ is 
a positive constant, whereas $\delta$ can be positive or negative.
The function, $\xi(\sigma)$, is written in terms of the Dedekind
$\eta$-function and is well approximated as $\xi(\sigma) \simeq
-(2\pi/3) {\rm cos}\,{\rm h}\sigma$. The dilatonic higher-order
loop and derivative corrections may be included as in
Ref.~\cite{Bru}, but we do not take them into account in this
work. In fact singularity-free cosmological solutions can be
constructed if the modulus-dependent term dominates over 
the dilatonic one in Eq.\,(\ref{Lc}).

With the flat Friedmann-Robertson-Walker (FRW) background
described by the scale factor $a$, 
the variation of the action (\ref{Bein}) yields the following background equations
\begin{eqnarray}
\label{back1} & & 12H^2=\dot{\phi}^2+2\dot{\chi}^2+3\dot{\sigma}^2
+2m_{\chi}^2\chi^2-96H^3\dot{f} \,, \\
& & 8(1+8H\dot{f})(\dot{H}+H^2) + 4(1+8\ddot{f})H^2\nonumber\\ &
&\,\,\,\,\,\, +\dot{\phi}^2 +2\dot{\chi}^2+3\dot{\sigma}^2
-2m_{\chi}^2\chi^2=0\,, \\
& & \ddot{\phi}+3H\dot{\phi}+2f_{\phi}
R_{\rm GB}^2=0\,, \\
& & \ddot{\sigma}+3H\dot{\sigma}-(2/3)f_{\sigma}
R_{\rm GB}^2=0\,, \\
& & \ddot{\chi}+3H\dot{\chi}+m_{\chi}^2\chi=0\,, \label{back}
\end{eqnarray}
where a dot denotes the derivative in terms of cosmic time $t$,  
$H \equiv \dot{a}/a$ is the Hubble parameter and
$f=(1/16)\left[\lambda
 e^{-\phi}-\delta\xi(\sigma) \right]$.
 The Gauss-Bonnet term is simply given as $R_{\rm
GB}^2=24H^2(\dot{H}+H^2)$.

The universe starts out from a weak coupling regime with $e^{\phi}
\ll 1$. In the absence of the modulus, the dynamics of the system
in the weak coupling region is characterized by the dilaton-driven
phase in PBB cosmology \cite{Veneziano:91}. During this phase the
universe contracts as $a \propto (-t)^{1/3}$ in the Einstein frame.
This solution does not connect to our
expanding branch in a nonsingular way in the tree-level action. We
can construct singularity-free bouncing solutions if higher-order
loop and derivative corrections are taken into account \cite{Bru}.
Nevertheless it is required to fine-tune the coefficients of the
correction terms. In addition the perturbations inside the Hubble
radius exhibit a negative instability \cite{Cartier}, which casts
doubt for the validity of linear perturbation theory.

The situation is different when the modulus field is present. If
the modulus dynamically controls the system rather than the
dilaton in Eqs.~(\ref{back1})-(\ref{back}), one gets the
super-inflationary solution with a growing Hubble rate in the
Einstein frame. In particular we can construct nonsingular
solutions by taking into account the correction (\ref{Lc}) for
negative $\delta$. For positive $\delta$, the first-order
$\alpha'$ corrections do not help to lead to a successful graceful
exit, as analyzed in Ref.~\cite{Toporensky}.

\begin{figure}
\includegraphics[scale=0.5]{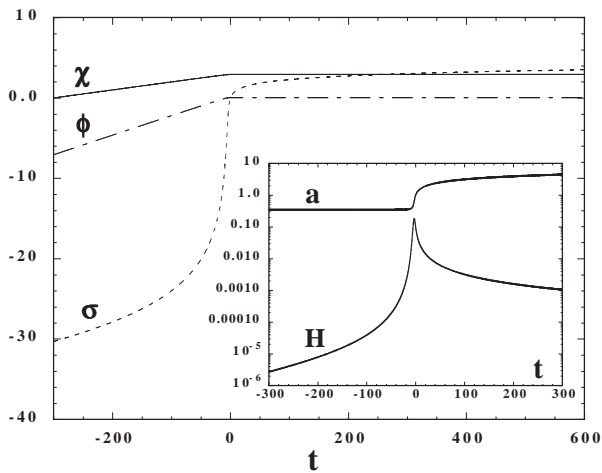}
\caption{The evolution of the dilaton $\phi$, the modulus
$\sigma$ and the inflaton $\chi$ before slow-roll inflation 
with initial conditions $H_i=2.71
\times 10^{-6}$, $\phi_i=-7.066$, $\dot{\phi}_i=0.024373$,
$\sigma_i=-30.31$, $\dot{\sigma}_i=0.021$, $\chi_i=0$ and
$\dot{\chi}_i=0.01$. The inflaton is kicked up along the potential
hill during the super-inflationary phase with a growing Hubble rate
($t<0$), which is followed by a phase of the frozen inflaton after
the graceful exit. {\bf Inset}: The evolution of the scale factor
$a$ and the Hubble rate $H$. } \label{evolution}
\end{figure}

When the modulus-type $\alpha'$ corrections dominate the system,
we get the background evolution for $t<0$ as \cite{Kawai}
\begin{eqnarray}
a \simeq a_0,~~~H=H_0/t^2,~~~\dot{\sigma}=5/t\,, \label{backmodu}
\end{eqnarray}
where $a_0$ and $H_0$ ($>0$) are constants. When one starts in an
expanding branch, the background is characterized by the
super-inflationary solution (\ref{backmodu}) until the graceful
exit ($t \simeq 0$). The Universe expands slowly initially  with a nearly
constant scale factor. Eventually the Hubble rate reaches its
peak value $H=H_{\rm max}$, after which the system connects to a
Friedmann-like branch. In the numerical simulation of
Fig.~\ref{evolution}, $H_{\rm max}$ corresponds to of order 0.1.
To ensure the validity of the theory, this maximum value should be
less than of order unity (i.e., less than the Planck energy). In
fact there is a negative instability for the perturbations if
$H_{\rm }$ is larger than unity \cite{Kawai}.

We include the dilaton in our simulations by generalizing
the analysis of Refs.~\cite{Antoniadis,Toporensky,Kawai}. 
As we find  in Fig.~\ref{evolution}, it is possible to obtain
singularity-free solutions even when the dilaton is present as
long as the modulus $\alpha'$ correction dominates over the
dilatonic one. If the super-inflationary modulus-driven phase
leads to a large number of $e$-foldings and also generates a
nearly scale-invariant density perturbation, this
can be an alternative to standard slow-roll inflation.
Unfortunately the scale factor evolves very slowly during this
modulus-driven phase (see Fig.~\ref{evolution}). In addition the
spectrum of the density perturbation generated in modulus-driven
inflation is highly blue-tilted, as we see later.

We are interested in whether the presence of the modulus-driven
phase can set the sufficient initial conditions for inflaton to
drive slow-roll inflation after the graceful exit. 
We shall consider a natural situation where the inflaton is
initially located around the potential minimum, $\chi \simeq 0$. If the
contribution of the potential term is dropped in Eq.~(\ref{back}),
one has $\dot{\chi} \propto a^{-3} \sim {\rm const}$. Therefore
$\chi$ grows linearly during the super-inflationary phase, whose
growth continues up to the Hubble peak [see Eq.~(\ref{backmodu})].
In Fig.~\ref{evolution} we show one example of the evolution for
$\chi$. The inflaton is driven up its potential hill during the
modulus-driven stage even in the presence of the inflaton
potential. The maximum value of $\chi$ ($=\chi_{\rm max}$) reached
at the Hubble peak is $\chi_{\rm max}=2.9$ in a Planck unit in
this case. This maximum value increases if the initial values of
$\dot{\chi}$ get larger.

We wish to analyze the case in which slow-roll inflation occurs
after the graceful exit. If the large-scale power spectrum
observed  in CMB is generated during this second stage of
inflation, the mass $m_{\chi}$ is constrained to be $m_{\chi}
\simeq 10^{-6}m_{\rm Pl}$ from the COBE normalization ($m_{\rm
Pl}$ is the Planck mass). Since this value is much smaller than
$m_{\rm Pl}$, the energy density of the inflaton is practically
negligible compared to that of the modulus around the graceful
exit when $H_{\rm max}$ is not too much smaller than unity. Therefore
the presence of the inflaton potential is not harmful for the
construction of nonsingular cosmological solutions.

The radiation is expected to be created by the decays of the
modulus and the dilaton after the transition to the Friedmann branch. 
It was shown in Ref.~\cite{Barreiro:1998aj} that the dilaton can be stabilized by
the potential generated by a gaugino condensation in the
presence of the radiation. We shall assume that the dilaton and
the modulus fall down to their potential minimums soon after the
graceful exit through a similar mechanism.

The inflaton freezes with a value $\chi \simeq \chi_{\rm max}$ due
to a large friction term until the Hubble rate drops down to of
order $m_{\chi}$. Since the energy density of the radiation
decreases faster than that of the field $\chi$, the background
dynamics is eventually dominated by the inflaton. The slow-roll
inflation takes place when the inflaton starts to roll down along
the potential for $H \lesssim m_{\chi}$. The number of
$e$-foldings is dependent on the field value $\chi_{\rm max}$.
Since we numerically found that $\chi_{\rm max}$ can exceed 3 in a Planck
unit, it is possible to have more than 60 $e$-foldings required to
solve horizon and flatness problems. We are interested in the
case where the information of the mudulus-driven phase before
the transition can be imprinted on the largest scales in CMB, 
corresponding to the value
of $\chi_{\rm max}$ that is slightly less than $3m_{\rm Pl}$.

\begin{figure}
\includegraphics[scale=0.45]{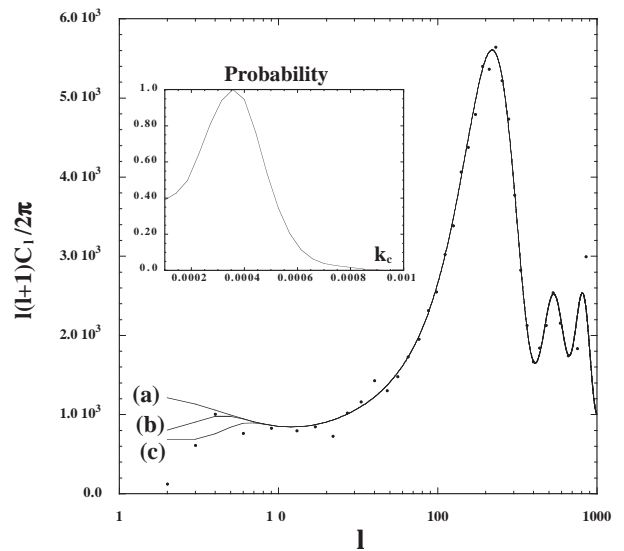}
\caption{ The CMB angular power spectrum showing the effects of
suppression of power at low multipoles. Curve~(a) corresponds to
the best-fit without a cut-off ($k_c=0$). Curve~(b) shows the best-fit
with a cut-off, corresponding to $k_c = 3.5
\times 10^{-4}\,{\rm Mpc}^{-1}$.
Curve~(c) is the case with 
$k_c=5.0 \times 10^{-4}\,{\rm Mpc}^{-1}$. 
{\bf Inset}: The
marginalized probability distribution for the cut-off $k_c$. }
\label{TT}
\end{figure}

Let us consider scalar perturbations about the FRW background
with a single-field, $\psi$. Note that our system involves multiple
scalar fields, but it is sufficient to study the perturbations of
one dynamically important field in estimating the spectrum
of metric perturbations. Then it is convenient to introduce a
gauge-invariant variable, 
\begin{eqnarray}
{\cal R} \equiv \vp-(H/\dot{\psi})\delta\psi\,, \label{metric}
\end{eqnarray}
where $\vp$ is the scalar perturbation in the perturbed 
metric \cite{Hwang:99}.
In the uniform-field gauge each Fourier mode of the perturbed Einstein 
equation with the correction (\ref{Lc}) is written 
as \cite{Kawai}
\begin{eqnarray}
\ddot{{\cal R}}_k+(2/t)\dot{{\cal R}}_k-\beta (k^2/a_0^2) t
\,{\cal R}_k=0\,, \label{dR_k}
\end{eqnarray}
where $k$ is a comoving wavenumber and $\beta$ and $a_0$ are
constants. The solution of this equation is written in
terms of the Bessel functions, $J_{\pm 1/3}$, and the spectrum of
the large-scale curvature perturbation is given by
\begin{eqnarray}
{\cal P}_{\cal R}\equiv k^3 \langle |{\cal R}_k|^2\rangle/2\pi^2
\propto k^{n_1-1}\,,~{\rm with}~ n_1=10/3\,, \label{spectral}
\end{eqnarray}
which is a highly blue-tilted spectrum \cite{Kawai}. Note that
when the kinematic term dominates over the potential term the
blue-tilted spectra are generic features even in generalized
Einstein theories \cite{Hwang:1996hz,Tsujikawa:01}. Around the
graceful exit, $\beta$ can be negative when the Hubble rate grows
larger than the Planck energy \cite{Kawai}, which means that
perturbations on the scales inside the Hubble radius exhibit a
negative instability. However, we can avoid this instability as
long as $H_{\rm max}$ is smaller than of order the Planck energy.
This is contrasted with the dilaton-driven inflation in which
inclusion of the $\alpha'$ correction is accompanied by a similar
negative instability even for $H_{\rm max} \lesssim m_{\rm Pl}$
\cite{Cartier}. Therefore perturbations in the modulus-driven
case which crossed the Hubble radius around the graceful exit is
more stable than in the dilaton-driven case.

The spectral index of curvature perturbations generated in 
slow-roll inflation after the graceful exit is close to
scale-invariant and is given as $n_2 \simeq 1-(1/\pi)(m_{\rm
Pl}/\chi)^2$ \cite{quantum}. When $\chi_{\rm max} \gtrsim 3m_{\rm
Pl}$, corresponding to the $e$-folds $N \gtrsim 60$, the CMB
spectrum on cosmologically relevant scales is determined by the
spectral index $n_2$. Meanwhile, when $\chi_{\rm max} \lesssim
3m_{\rm pl}$, the spectrum (\ref{spectral}) is within an
observational range on the scales larger than a cut-off which
connects two power spectra. Then we simply model the power
spectrum to be ${\cal P}_{\cal R}=A_1 (k/k_0)^{n_1-1}$ for $k \le
k_c$ and ${\cal P}_{\cal R}=A_2 (k/k_0)^{n_2-1}$ for $k \ge k_c$,
where $k_c$ is a cut-off wavenumber. As shown in Fig.~\ref{TT}
it is possible to explain the loss of power at low multipoles by
employing the highly blue-tilted spectrum (\ref{spectral}), as
long as $\chi_{\rm max}$ is slightly less than $3m_{\rm Pl}$. We
have carried out the likelihood analysis by varying 7 cosmological
\& inflationary parameters using the recent WMAP \cite{Spergel}
and 2dF \cite{2dF} data sets, and found that the best-fit value of
the cut-off scale corresponds to $k_c = 3.5 \times 10^{-4}\,{\rm
Mpc}^{-1}$ (see the inset of Fig.\,\ref{TT}). The smaller cut-off
corresponding to $\chi_{\rm max} \gtrsim 3m_{\rm Pl}$ is within the
$2\sigma$ contour bound, thus not ruled out observationally. 
We also analyzed the smoothly connected power spectrum, 
$P_{\cal R}=A(k/k_0)^{n_2-1} \left[1-\exp \{-(k/k_c)^{n_1-n_2}
 \}\right]$, around the cut-off $k_c$, which may be a possible case 
when higher-order loop corrections become important around the
graceful exit.
We obtained a similar likelihood value of the cut-off as 
shown in Fig.\,2, which is also consistent with the
past related works in Refs.~\cite{sup,Bridle}.

In summary, we studied a nonsingular cosmological scenario
based on the orbifold compactification in low-energy effective
string theory and showed that the inflaton is driven up
the potential hill during the modulus-driven stage before the
graceful exit, thus setting up sufficient initial conditions for
following slow-roll inflation. The presence of the modulus-driven
phase generates a blue-tilted primordial power spectrum, which can
explain the loss of power observed in CMB low multipoles. 
The spectral index (\ref{spectral}) is much larger than in
the case of noncommutative inflation \cite{noncom} or loop
quantum gravity \cite{quantum}, thus leading to a stronger
suppression. Our work provides an exciting possibility to pick up
the signature of extra dimensions from  high-precision observations.

{\bf Acknowledgements}~ We thank Bo Feng, Burin Gumjudpai, Roy
Maartens, Parampreet Singh for useful discussions and Antony
Lewis, David Parkinson for kind support in interpreting the
likelihood analysis. S.T. is thankful for financial support from
JSPS (No.\,04942). The work of Y.P. and X.Z. is supported by the
NSF of China and also by the Ministry of Sceince and Technology of
China under grant No. NKBRSF G19990754.


\end{document}